\documentclass[english]{scrartcl}

\usepackage{fullpage} 
\usepackage{amsfonts}
\usepackage[dvipdfm]{hyperref} 
\usepackage{graphicx}

\newcommand{\expt}{{\rm E}}
\newcommand{\var}{{\rm Var}}
\newcommand{\cov}{{\rm Cov}}

\title{EM algorithm and variants: an informal tutorial}

\author{
Alexis Roche\thanks{alexis.roche@gmail.com} \\
Service Hospitalier Fr\'ed\'eric Joliot, CEA, F-91401 Orsay, France
}

\date{Spring 2003 (revised: September 2012)}

\begin{document}
\maketitle

\section{Introduction}

The expectation-maximization (EM) algorithm introduced by Dempster et
al~\cite{Dempster-77} in 1977 is a very general method to solve
maximum likelihood estimation problems. In this informal report, we
review the theory behind~EM as well as a number of EM~variants,
suggesting that beyond the current state of the art is an even much
wider territory still to be discovered.

\section{EM background}
\label{sec:em}

Let~$Y$ a random variable with probability density function~(pdf)
$p(y|\theta)$, where~$\theta$ is an unknown parameter vector. Given an
outcome~$y$ of~$Y$, we aim at maximizing the likelihood function
${\cal L}(\theta) \equiv p(y|\theta)$ wrt~$\theta$ over a given search
space $\Theta$. This is the very principle of maximum likelihood (ML)
estimation. Unfortunately, except in not very exciting situations such
as, e.g. estimating the mean and variance of a Gaussian population, a
ML~estimation problem has generally no closed-form solution. Numerical
routines are then needed to approximate it.

\subsection{EM as a likelihood maximizer}

The EM~algorithm is a class of optimizers specifically taylored to ML
problems, which makes it both general and not so general.  Perhaps the
most salient feature of EM is that it works iteratively by maximizing
successive local approximations of the likelihood function. Therefore,
each iteration consists of two steps: one that performs the
approximation (the E-step) and one that maximizes it (the
M-step). But, let's make it clear, not any two-step iterative scheme
is an EM~algorithm. For instance, Newton and quasi-Newton
methods~\cite{Press-92} work in a similar iterative fashion but do not
have much to do with~EM. What essentially defines an EM~algorithm is
the philosophy underlying the local approximation scheme -- which, in
particular, doesn't rely on differential calculus.

The key idea underlying~EM is to introduce a latent variable~$Z$ whose
pdf depends on~$\theta$ with the property that maximizing
$p(z|\theta)$ is easy or, say, easier than maximizing
$p(y|\theta)$. Loosely speaking, we somehow enhance the incomplete
data by guessing some useful additional information.  Technically,
$Z$~can be any variable such that~$\theta\to Z\to Y$ is a Markov
chain\footnote{In many presentations of~EM, $Z$~is as an aggregate
variable~$(X,Y)$, where~$X$ is some ``missing'' data, which
corresponds to the special case where the transition~$Z\to Y$ is
deterministic. We believe this restriction, although important in
practice, is not useful to the global understanding of~EM. By the way,
further generalizations will be considered later in this report.},
i.e. we assume that~$p(y|z,\theta)$ is independent from~$\theta$,
yielding a Chapman-Kolmogorov equation:
\begin{equation}
\label{eq:complete_data_space}
p(z,y|\theta)=p(z|\theta)p(y|z)
\end{equation}

Reasons for that definition will arise soon. Conceptually, $Z$~is a
complete-data space in the sense that, if it were fully observed, then
estimating~$\theta$ would be an easy game. We will emphasize that the
convergence speed of~EM is highly dependent upon the complete-data
specification, which is widely arbitrary despite some estimation
problems may have seamingly ``natural'' hidden variables. But, for the
time being, we assume that the complete-data specification step has
been accomplished.

\subsection{EM as a consequence of Jensen's inequality}
\label{sec:jensen}

Quite surprisingly, the original EM~formulation stems from a very
simple variational argument. Under almost no assumption regarding the
complete variable~$Z$, except its pdf doesn't vanish to zero, we can
bound the variation of the log-likelihood function
$L(\theta)\equiv\log p(y|\theta)$ as follows:
\begin{eqnarray}
L (\theta) - L (\theta') 
  & = & \log \frac{p(y|\theta)}{p(y|\theta')} \nonumber \\
  & = & \log \int \frac{p(z,y|\theta)}{p(y|\theta')} \, dz \nonumber \\
  & = & \log \int \frac{p(z,y|\theta)}{p(z,y|\theta')} \, p(z|y,\theta')\, dz \nonumber\\
  & = & \log \int \frac{p(z|\theta)}{p(z|\theta')} \, p(z|y,\theta')\, dz 
  \label{eq:markov}\\
  & \geq & \underbrace{\int \log \frac{p(z|\theta)}{p(z|\theta')}\,p(z|y,\theta')\,dz}_{
    {\rm Call\ this\ }Q(\theta,\theta')}
  \label{eq:jensen} 
\end{eqnarray}

Step~(\ref{eq:markov}) results from the fact that~$p(y|z,\theta)$ is
independent from~$\theta$ owing to~(\ref{eq:complete_data_space}).
Step~(\ref{eq:jensen}) follows from Jensen's inequality
(see~\cite{Cover-91} and appendix~\ref{app:jensen}) along with the
well-known concavity property of the logarithm function. Therefore,
$Q(\theta,\theta')$ is an auxiliary function for the log-likelihood,
in the sense that: {\em (i)}~the likelihood variation from $\theta'$
to $\theta$ is always greater than $Q(\theta,\theta')$, and {\em
(ii)}~we have $Q(\theta',\theta')=0$. Hence, starting from an initial
guess $\theta'$, we are guaranteed to increase the likelihood value if
we can find a~$\theta$ such that $Q(\theta,\theta')>0$. Iterating such
a process defines an EM algorithm.

There is no general convergence theorem for~EM, but thanks to the
above mentioned monotonicity property, convergence results may be
proved under mild regularity conditions. Typically, convergence
towards a non-global likelihood maximizer, or a saddle point, is a
worst-case scenario. Still, the only trick behind EM is to exploit the
concavity of the logarithm function!

% EM is a direct consequence of Jensen's inequality!

\subsection{EM as expecation-maximization}
\label{sec:classical}

Let's introduce some notations. Developing the logarithm in the
right-hand side of~(\ref{eq:jensen}), we may interpret our auxiliary
function as a difference: $Q(\theta,\theta') = Q(\theta|\theta') -
Q(\theta'|\theta')$, with:
\begin{equation}
\label{eq:auxiliary}
Q(\theta|\theta') \equiv \int \log
p(z|\theta)\,p(z|y,\theta')\,dx
\end{equation}

Clearly, for a fixed~$\theta'$, maximizing $Q(\theta,\theta')$ wrt
$\theta$ is equivalent to maximizing $Q(\theta|\theta')$.  If we
consider the residual function: $R(\theta|\theta')\equiv
L(\theta)-Q(\theta|\theta')$, the incomplete-data log-likelihood may
be written as:
$$ 
L(\theta) = Q(\theta|\theta') + R(\theta|\theta')
$$

The EM~algorithm's basic principle is to replace the maximization
of~$L(\theta)$ with that of~$Q(\theta|\theta')$, hopefully easier to
deal with. We can ignore $R(\theta|\theta')$ because
inequality~(\ref{eq:jensen}) implies that $R(\theta|\theta') \geq
R(\theta'|\theta')$. In other words, EM~works because the auxiliary
function $Q(\theta|\theta')$ always deteriorates as a likelihood
approximation when~$\theta$ departs from~$\theta'$. In an ideal world,
the approximation error would be constant; then, maximizing~$Q$ would,
not only increase, but truly maximize the likelihood. Unfortunately,
this won't be the case in general. Therefore, unless we decide to give
up on maximizing the likelihood, we have to iterate -- which gives
rise to quite a popular statistical learning algorithm.

Given a current parameter estimate $\theta_n$:
\begin{itemize}

\item E-step: form the auxiliary function $Q(\theta|\theta_n)$ as
defined in~(\ref{eq:auxiliary}), which involves computing the
posterior distribution of the unobserved variable, $p(z|y,\theta_n)$.
The ``E'' in E-step stands for ``expectation'' for reasons that will
arise in section~\ref{sec:conditional_expectation}.

\item M-step: update the parameter estimate by maximizing the
auxiliary function:
$$
\theta_{n+1} = \arg\max_\theta Q(\theta|\theta_n)
$$

An obvious but important generalization of the M-step is to replace
the maximization with a mere increase of $Q(\theta|\theta_n)$. Since,
anyway, the likelihood won't be maximized in one iteration, increasing
the auxiliary function is enough to ensure that the likelihood will
increase in turn, thus preserving the monotonicity property
of~EM. This defines generalized EM (GEM) algorithms. More on this
later.
\end{itemize}

\subsection{Some probabilistic interpretations here...}
\label{sec:conditional_expectation}

For those familiar with probability theory, $Q(\theta|\theta')$ as
defined in~(\ref{eq:auxiliary}) is nothing but the conditional
expectation of the complete-data log-likelihood in terms of the
observed variable, taken at~$Y=y$, and assuming the true parameter
value is~$\theta'$:
\begin{equation}
\label{eq:cecl}
Q(\theta|\theta') \equiv \expt \big[\log p(Z|\theta)|y,\theta'\big]
\end{equation}

This remark explains the ``E'' in E-step, but also yields some
probabilistic insight on the auxiliary function.  For all~$\theta$,
$Q(\theta|\theta')$ is an estimate of the the complete-data
log-likelihood that is built upon the knowledge of the incomplete data
and under the working assumption that the true parameter value is
known. In some way, it is not far from being the ``best'' estimate
that we can possibly make without knowing~$Z$, because conditional
expectation is, by definition, the estimator that minimizes the
conditional mean squared error\footnote{For all $\theta$, we have:
  $\displaystyle Q(\theta|\theta') = \arg\min_{\mu} \ \int \big[ \log
    p(z|\theta) - \mu \big]^2 \, p(z|y,\theta')\,dz$.}.

Having said that, we might still be a bit suspiscious. While we can
grant that $Q(\theta|\theta')$ is a reasonable estimate of the
complete-data log-likelihood, recall that our initial problem is to
maximize the {\em incomplete-data} (log) likelihood. How good a fit is
$Q(\theta|\theta')$ for $L(\theta)$? To answer that, let's see a bit
more how the residual~$R(\theta|\theta')$ may be interpreted. We have:
\begin{eqnarray}
R(\theta|\theta')
  & = & \log p(y|\theta) - \int \log p(z|\theta)\,p(z|y,\theta')\,dz \nonumber\\
  & = & \int \log \frac{p(y|\theta)}{p(z|\theta)}\,p(z|y,\theta')\,dz \nonumber\\
  & = & \int \log \frac{p(y|z,\theta)}{p(z|y,\theta)}\,p(z|y,\theta')\,dz,
\end{eqnarray}
where the last step relies on Bayes' law. Now, $p(y|z,\theta)=p(y|z)$
is independent from~$\theta$ by the Markov
property~(\ref{eq:complete_data_space}).  Therefore, using the
simplified notations $q_{\theta}(z)\equiv p(z|y,\theta)$ and
$q_{\theta'}(z) \equiv p(z|y,\theta')$, we get:
\begin{equation}
\label{eq:kullback}
R(\theta|\theta') - R(\theta'|\theta')
=
\underbrace{\int \log \frac{q_{\theta'}(z)}{q_{\theta}(z)}
\, q_{\theta'}(z)\,dz}_{{\rm Call\ this\ } D(q_{\theta'}\|q_{\theta})}
\end{equation}

In the language of information theory, this quantity
$D(q_{\theta'}\|q_{\theta})$ is known as the Kullback-Leibler
distance, a general tool to assess the deviation between two
pdfs~\cite{Cover-91}. Although it is not, strictly speaking, a genuine
mathematical distance, it is always positive and vanishes iff the pdfs
are equal which, again and not surprinsingly, comes as a direct
consequence of Jensen's inequality.

What does that mean in our case? We noticed earlier that the
likelihood approximation~$Q(\theta|\theta')$ cannot get any better
as~$\theta$ deviates from~$\theta'$. We now realize from
equation~(\ref{eq:kullback}) that this property reflects an implicit
strategy of ignoring the variations of~$p(z|y,\theta)$
wrt~$\theta$. Hence, a perfect approximation would be one for
which~$p(z|y,\theta)$ is independent from~$\theta$. In other words, we
would like $\theta\to Y\to Z$ to define a Markov chain... But, look,
we already assumed that $\theta\to Z\to Y$ is a Markov chain. Does the
Markov property still hold when permuting the roles of~$Y$ and~$Z$?

From the fundamental data processing inequality~\cite{Cover-91}, the
answer is no in general. Details are unnecessary here. Just remember
that the validity domain of~$Q(\theta|\theta')$ as a local likelihood
approximation is controlled by the amount of information that both~$y$
and~$\theta$ contain about the complete data. We are now going to
study this aspect more carefully.

\subsection{EM as a fix point algorithm and local convergence}
\label{sec:fixpoint}

Quite clearly, EM is a fix point algorithm:
$$
\theta_{n+1} = \Phi(\theta_n)
\qquad
{\rm with}\quad
\Phi(\theta') = \arg\max_{\theta\in\Theta} Q(\theta,\theta')
$$

Assume the sequence $\theta_n$ converges towards some value
$\hat{\theta}$ -- hopefully, the maximum likelihood estimate but
possibly some other local maximum or saddle point. Under the
assumption that $\Phi$ is continuous, $\hat{\theta}$ must be a fix
point for $\Phi$, i.e.  $\hat{\theta}=\Phi(\hat{\theta})$.
Furthermore, we can approximate the sequence's asymptotic behavior
using a first-order Taylor expansion of~$\Phi$ around $\hat{\theta}$,
which leads to:
$$
\theta_{n+1} 
\approx
S \hat{\theta}
+ 
(I-S) \theta_n
\qquad {\rm with}\quad
S = I - \frac{\partial \Phi}{\partial \theta}\big|_{\hat{\theta}}
$$ 

This expression shows that the rate of convergence is controlled
by~$S$, a square matrix that is constant across iterations. Hence,
$S$~is called the speed matrix, and its spectral radius\footnote{Let
$(\lambda_1,\lambda_2,\ldots,\lambda_m)$ be the complex eigenvalues
of~$S$. The spectral radius is $\rho(S)=\min_i |\lambda_i|$.}  defines
the global speed. Unless the global speed is one, the local
convergence of~EM is only linear. We may relate~$S$ to the likelihood
function by exploiting the fact that, under sufficient smoothness
assumptions, the maximization of~$Q$ is characterized by:
$$
\frac{\partial Q}{\partial \theta^t} (\theta_{n+1},\theta_n)
=0
$$ 

From the implicit function theorem, we get the gradient of~$\Phi$:
$$
\frac{\partial \Phi}{\partial \theta}
=
- \Big(\frac{\partial^2 Q}{\partial \theta \partial \theta^t}\Big)^{-1}
\frac{\partial^2 Q}{\partial \theta' \partial \theta^t}
\quad
\Rightarrow
\quad
S = 
\Big(\frac{\partial^2 Q}{\partial \theta \partial \theta^t}\Big)^{-1}
\Big[
\frac{\partial^2 Q}{\partial \theta \partial \theta^t}
+ \frac{\partial^2 Q}{\partial \theta' \partial \theta^t}
\Big]
$$
where, after some manipulations:
\begin{eqnarray*}
\frac{\partial^2 Q}
{\partial \theta \partial \theta^t}\big|_{(\hat{\theta},\hat{\theta})} 
       & = & 
       \int p(z|y,\hat{\theta})\,
       \underbrace{ \frac{\partial^2\log p(z|\theta)}
       {\partial \theta \partial \theta^t}\big|_{\hat{\theta}}
       }_{ {\rm Call\ this\ } - J_z(\hat{\theta}) }\, dz \\[1em]
\frac{\partial^2 Q}
{\partial \theta' \partial \theta^t}\big|_{(\hat{\theta},\hat{\theta})}
       & = & 
       \underbrace{
       \frac{\partial^2\log p(y|\theta)}
       {\partial \theta \partial \theta^t}\big|_{\hat{\theta}}
       }_{ {\rm Call\ this\ } -  J_y(\hat{\theta}) } 
       \ - \ \frac{\partial^2 Q}
       {\partial \theta \partial \theta^t}\big|_{(\hat{\theta},\hat{\theta})} 
\end{eqnarray*}

The two quantities $J_y(\hat{\theta})$ and $J_z(\hat{\theta})$ turn
out to be respectively the observed-data information matrix and the
complete-data information matrix.  The speed matrix is thus given by:
\begin{equation}
\label{eq:speed_matrix}
S = 
{\cal J}_z(\hat{\theta})^{-1}
J_y(\hat{\theta})
\qquad
{\rm with}\quad
{\cal J}_z(\hat{\theta})
\equiv \expt\big[J_z(\hat{\theta})|y,\hat{\theta}\big]
\end{equation}

We easily check that: ${\cal J}_z(\hat{\theta})= J_y(\hat{\theta}) +
{\cal F}_{z|y}(\hat{\theta})$, where ${\cal F}_{z|y}(\hat{\theta})$ is
the Fisher information matrix corresponding to the posterior
pdf~$p(z|y, \hat{\theta})$, which is always symmetric
positive. Therefore, we have the alternative expression:
$$
S = 
\big[ 
J_y(\hat{\theta}) + {\cal F}_{z|y}(\hat{\theta}) 
\big]^{-1} 
J_y(\hat{\theta})
$$

For fast convergence, we want~$S$ close to identity, so we better have
the posterior Fisher matrix as ``small'' as possible.  To interpret
this result, let's imagine that~$Z$ is drawn from
$p(z|y,\hat{\theta})$, which is not exactly true since $\hat{\theta}$
may be at least slightly different from the actual parameter
value. The Fisher information matrix represents the average
information that the complete data contains about $\hat{\theta}$
conditional to the observed data. In this context, ${\cal
F}_{z|y}(\hat{\theta})$ is a measure of missing information, and the
speed matrix is the fraction of missing data. The conclusion is that
the rate of convergence of EM is governed by the fraction of missing
data.

\subsection{EM as a proximal point algorithm}
\label{sec:proximal}

Chr\'etien \& Hero \cite{Chretien-00} note that EM~may also be
interpreted as a proximal point algorithm, i.e. an iterative scheme of
the form:
\begin{equation}
\label{eq:proximal}
\theta_{n+1}
=
\arg\max_{\theta}
\big[
L(\theta) - \lambda_n \Psi( \theta, \theta_n )
\big],
\end{equation}
where $\Psi$ is some positive regularization function and $\lambda_n$
is a sequence of positive numbers.

Let us see where this result comes from. In
section~\ref{sec:conditional_expectation}, we have established the
fundamental log-likelihood decomposition underlying~EM, $L(\theta) =
Q(\theta|\theta') + R(\theta|\theta')$, and related the variation of
$R(\theta|\theta')$ to a Kullback distance~(\ref{eq:kullback}). Thus,
for some current estimate $\theta_n$, we can write:
$$
Q(\theta|\theta_n) 
= L(\theta) - D(q_{\theta_n}\|q_{\theta})
- R(\theta_n|\theta_n),
$$ where $q_{\theta}(z)\equiv p(z|y,\theta)$ and $q_{\theta_n}(z)
\equiv p(x|y,\theta_n)$ are the posterior pdfs of the complete data,
under $\theta$ and $\theta_n$, respectively. From this equation, it
becomes clear that maximizing $Q(\theta|\theta_n)$ is equivalent to an
update rule of the form~(\ref{eq:proximal}) with:
$$
\Psi( \theta, \theta_n )
= 
D(q_{\theta_n}\|q_{\theta}),
\qquad 
\lambda_n \equiv 1
$$

The proximal interpretation of~EM is very useful to derive general
convergence results \cite{Chretien-00}. In particular, the convergence
rate may be superlinear if the sequence $\lambda_n$ is chosen so as to
converge towards zero. Unfortunately, such generalizations are usually
intractable because the objective function may no longer simplify as
soon as $\lambda_n\not= 1$.

\subsection{EM as maximization-maximization}
\label{sec:global}

Another powerful way of conceptualizing~EM is to reinterpret the
E-step as another maximization. This idea, which was formalized only
recently by Neal \& Hinton \cite{Neal-98}, appears as a breakthrough
in the general understanding of EM-type procedures. Let us consider
the following function:
\begin{equation}
\label{eq:maxmax}
L(\theta,q) 
\equiv \expt_{q} \big[\log p(Z,y|\theta)\big] + H(q)
\end{equation}
where $q(z)$ is some pdf (yes, any pdf), and $H(q)$ is its entropy
\cite{Cover-91}, i.e. $H(q)\equiv - \int \log q(z)\,q(z)\,dz$. We
easily obtain an equivalent expression that involves a Kullback-Leiber
distance:
$$
L(\theta,q) = 
L(\theta) - D(q\|q_{\theta}),
$$ where we still define $q_{\theta}(z)\equiv p(z|y,\theta)$ for
notational convenience. The last equation reminds us immediately of
the proximal interpretation of~EM which was briefly discussed in
section~\ref{sec:proximal}.  The main difference here is that we don't
impose~$q(z)=p(z|y,\theta)$ for some~$\theta$. Equality holds for {\em
any} distribution!

Assume we have an initial guess $\theta_n$ and try to find~$q$ that
maximizes $L(\theta_n,q)$. From the above discussed properties of the
Kullback-Leibler distance, the answer is~$q(z)=q_{\theta_n}(z)$. Now,
substitute $q_{\theta_n}$ in~(\ref{eq:maxmax}), and maximize
over~$\theta$: this is the same as performing a standard
M-step\footnote{To see that, just remember that
$p(z,y|\theta)=p(z|\theta)p(y|z)$ where $p(y|z)$ is independent from
$\theta$ due to the Markov property~(\ref{eq:complete_data_space}).}!
Hence, the conventional~EM algorithm boils down to an alternate
maximization of $L(\theta,q)$ over a search space $\Theta \times {\cal
Q}$, where ${\cal Q}$ is a suitable set of pdfs, i.e. ${\cal Q}$ must
include all pdfs from the set
$\{q(z)=p(z|y,\theta),\,\theta\in\Theta\}$. It is easy to check that
any global maximizer~$(\hat{\theta},\hat{q})$
of~$L(\hat{\theta},\hat{q})$ is such that $\hat{\theta}$ is also a
global maximizer of~$L(\theta)$. By the way, this is also true for
local maximizers under weak assumptions~\cite{Neal-98}.

The key observation of Neal \& Hinton is that the alternate scheme
underlying~EM may be replaced with other maximization strategies
without hampering the simplicity of~EM. In the conventional~EM
setting, the auxiliary pdf~$q_n(z)$ is always constrained to a
specific form. This is to say that~EM authorizes only specific
pathways in the expanded search space $\Theta \times {\cal Q}$,
yielding some kind of ``labyrinth'' motion. Easier techniques to find
its way in a labyrinth include breaking the walls or escaping through
the roof. Similarly, one may consider relaxing the maximization
constraint in the E-step. This leads for instance to incremental and
sparse EM~variants (see section~\ref{sec:deterministic_variants}).

\section{Deterministic EM variants}
\label{sec:deterministic_variants}

We first present deterministic EM~variants as opposed to stochastic
variants.  Most of these deterministic variants attempt at speeding up
the algorithm, either by simplifying computations, or by increasing
the rate of convergence (see section~\ref{sec:fixpoint}).

\subsection{CEM}
\label{sec:cem}

Classification~EM \cite{Celeux-92}. The whole~EM story is about
introducing a latent variable~$Z$ and performing some inference about
its posterior pdf. We might wonder: why not simply estimate~$Z$?  This
is actually the idea underlying the CEM~algorithm, which is a simple
alternate maximization of the functional $p(z,y|\theta)$ wrt
both~$\theta$ and~$z$. Given a current parameter estimate $\theta_n$,
this leads to:
\begin{itemize}
\item Classification step: find $\displaystyle z_n = \arg\max_z p(z|y,\theta_n)$. 
\item Maximization step: find $\displaystyle \theta_{n+1} = \arg\max_\theta p(z_n|\theta)$. 
\end{itemize}

Notice that a special instanciation of~CEM is the well-known $k$-means
algorithm. In practice, CEM~has several advantages over~EM, like being
easier to implement and typically faster to converge. However,
CEM~doesn't maximize the incomplete-data likelihood and, therefore,
the monotonicity property of~EM is lost. While CEM~estimates the
complete data explicitely, EM~estimates only sufficient statistics for
the complete data. In this regard, EM~may be understood as a fuzzy
classifier that avoids the statistical efficiency problems inherent to
the~CEM approach. Yet, CEM~is often useful in practice.

\subsection{Aitken's acceleration}
\label{sec:aitken}

An early EM~extension
\cite{Dempster-77,Louis-82,Meilijson-89}. Aitken's acceleration is a
general purpose technique to speed up the convergence of a fixed point
recursion with asymptotic linear behavior. Section~\ref{sec:fixpoint}
established that, under appropriate smoothness assumptions, EM~may be
approximated by a recursion of the form:
$$
\theta_{n+1} 
\approx
S \hat{\theta}
+ 
(I-S) \theta_n,
$$ where $\hat{\theta}$ is the unknown limit and~$S$ is the speed
matrix given by~(\ref{eq:speed_matrix}) which depends on this
limit. Aitken's acceleration stems from the remark that, if~$S$ was
known, then the limit could be computed explicitely in a single
iteration, namely: $ \hat{\theta} \approx \theta_0 + S^{-1}
(\theta_1-\theta_0)$ for some starting value~$\theta_0$. Despite
that~$S$ is unknown and the sequence is not strictly linear, we are
tempted to consider the following modified EM~scheme.  Given a current
parameter estimate~$\theta_n$,
\begin{itemize}
\item E-step: compute $Q(\theta|\theta_n)$ and approximate the inverse
speed matrix: $ \displaystyle S_n^{-1} = J_y(\theta_n)^{-1} \, {\cal
J}_z (\theta_n)$.
\item M-step: unchanged, get an intermediate value 
$\displaystyle \theta^* = \arg\max_\theta Q(\theta|\theta_n)$.
\item Acceleration step: update the parameter using
$\displaystyle \theta_{n+1} = \theta_n + S_n^{-1}(\theta^* - \theta_n)$.
\end{itemize}

It turns out that this scheme is nothing but the Newton-Raphson method
to find a zero of $\theta \mapsto \Phi(\theta)-\theta$, where~$\Phi$
is the map defined by the EM~sequence, i.e. $\Phi(\theta')=
\arg\max_\theta Q(\theta|\theta')$. Since the standard~EM sets $S_n=I$
on each iteration, it may be viewed as a first-order approach to the
same zero-crossing problem, hence avoiding the expense of
computing~$S_n$. Beside this important implementational issue,
convergence is problematic using Aitken's acceleration as the
monotonicity property of~EM is generally lost.

\subsection{AEM}
\label{sec:aem}

Accelerated~EM \cite{Jamshidian-93}. To trade off between~EM and its
Aitken's accelerated version (see section~\ref{sec:aitken}),
Jamshidian and Jennrich propose a conjugate gradient approach. Don't
be messed up: this is not a traditional gradient-based method
(otherwise there would be no point to talk about it in this
report). No gradient computation is actually involved in here.  The
``gradient'' is the function $\Phi(\theta)-\theta$, which may be
viewed as a generalized gradient for the incomplete-data
log-likelihood, hence justifying the use of the generalized conjugate
gradient method (see e.g.~\cite{Press-92}). Compared to the Aitken's
accelerated~EM, the resulting AEM~algorithm doesn't require computing
the speed matrix. Instead, the parameter update rule is of the form:
$$
\theta_{n+1} = \theta_n + \lambda_n d_n,
$$ where~$d_n$ is a direction composed from the current direction
$\Phi(\theta_n)-\theta_n$ and the previous directions (the essence of
conjugate gradient), and~$\lambda_n$ is a step size typically computed
from a line maximization of the complete-data likelihood (which may or
may not be cumbersome). As an advantage of line maximizations, the
monotonicity property of~EM is safe. Also, from this generalized
gradient perspective, it is straightforward to devise EM~extensions
that make use of other gradient descent techniques such as the
steepest descent or quasi-Newton methods~\cite{Press-92}.

\subsection{ECM}
\label{sec:ecm}

Expectation Conditional Maximization \cite{Meng-93}. This variant (not
to be confused with CEM, see above) was introduced to cope with
situations where the standard M-step is intractable. It is the first
on a list of coordinate ascent-based EM~extensions.

In~ECM, the M-step is replaced with a number of lower dimensional
maximization problems called CM-steps. This implies decomposing the
parameter space as a sum of subspaces, which, up to some possible
reparameterization, is the same as splitting the parameter vector into
several blocks, $\theta = (t_1, t_2, \ldots, t_s)$. Starting from a
current estimate $\theta_n$, the CM-steps update one coordinate block
after another by partially maximizing the auxiliary $Q$-function,
yielding a scheme similar in essence to Powell's multidimensional
optimization method \cite{Press-92}. This produces a sequence
$\theta_n = \theta_{n,0} \to \theta_{n,1} \to \theta_{n,2} \to \ldots
\to \theta_{n,s-1} \to \theta_{n,s} = \theta_{n+1}$, such that:
$$
Q(\theta_n|\theta_n)
\leq Q(\theta_{n,1}|\theta_n)
\leq Q(\theta_{n,2}|\theta_n)
\leq
\ldots
\leq Q(\theta_{n,s-1}|\theta_n)
\leq Q(\theta_{n+1}|\theta_n)
$$

Therefore, the auxiliary function is guaranteed to increase on each
CM-step, hence globally in the M-step, and so does the incomplete-data
likelihood. Hence, ECM is a special case of GEM (see
section~\ref{sec:classical}).

\subsection{ECME}
\label{sec:ecme}

ECM either \cite{Liu-94}. This is an extension of ECM where some
CM-steps are replaced with steps that maximize, or increase, the
incomplete-data log-likelihood $L(\theta)$ rather than the auxiliary
$Q$-function. To make sure that the likelihood function increases
globally in the M-step, the only requirement is that the CM-steps that
act on the actual log-likelihood be performed after the usual
$Q$-maximizations. This is because increasing the $Q$-function only
increases likelihood from the starting point, namely $\theta_n$, which
is held fixed during the M-step (at least, this is what we
assume)\footnote{ For example, if one chooses~$\theta^*$ such that
$L(\theta^*)\geq L(\theta_n)$ and, then, $\theta_{n+1}$ such that
$Q(\theta_{n+1}|\theta_n)\geq Q(\theta^*|\theta_n)$, the only
conlusion is that the likelihood increases from~$\theta_n$
to~$\theta^*$, but may actually decrease from~$\theta^*$
to~$\theta_{n+1}$ because $\theta^*$ is not the starting point
of~$Q$. Permuting the $L$-maximization and the $Q$-maximization, we
have $Q(\theta^*|\theta_n)\geq Q(\theta_n|\theta_n)$, thus
$L(\theta^*)\geq L(\theta_n)$, and therefore $L(\theta_{n+1})\geq
L(\theta_n)$ since we have assumed $L(\theta_{n+1})\geq
L(\theta^*)$. This argument generalizes easily to any intermediate
sequence using the same cascade inequalities as in the derivation
of~ECM (see section~\ref{sec:ecm}).}.

Starting with $Q$-maximizations is guaranteed to increase the
likelihood, and of course subsequent likelihood maximizations can only
improve the situation.  With the correct setting, ECME is even more
general than~GEM as defined in section~\ref{sec:classical} while
inheriting its fundamental monotonicity property. An example
application of ECME is in mixture models, where typically mixing
proportions are updated using a one-step Newton-Raphson gradient
descent on the incomplete-data likelihood, leading to a simple
additive correction to the usual EM~update rule \cite{Liu-94}. At
least in this case, ECME has proved to converge faster than
standard~EM.

\subsection{SAGE}
\label{sec:sage}

Space-Alternating Generalized EM \cite{Fessler-94,Fessler-95}. In the
continuity of ECM and ECME (see sections~\ref{sec:ecm} and
~\ref{sec:ecme}), one can imagine defining an auxiliary function
specific to each coordinate block of the parameter vector. More
technically, using a block decomposition
$\theta=(t_1,t_2,\ldots,t_s)$, we assume that, for each
block~$i=1,\ldots,s$, there exists a function $Q_i(\theta|\theta')$
such that, for all~$\theta$ and~$\theta'$ with identical block
coordinates except (maybe) for the $i$-th block, we have: $L(\theta) -
L(\theta') \geq Q_i(\theta|\theta') - Q_i(\theta'|\theta')$.

This idea has two important implications. First, the usual ECM scheme
needs to be rephrased, because changing the auxiliary function across
CM-steps may well result in decreasing the likelihood, a problem
worked around in~ECME with an appropriate ordering of the CM-steps. In
this more general framework, though, there may be no such fix to save
the day. In order to ensure monotonicity, at least some CM-steps
should start with ``reinitializing'' their corresponding auxiliary
function, which means...  performing an E-step. It is important to
realize that, because the auxiliary function is coordinate-specific,
so is the E-step. Hence, each ``CM-step'' becomes an EM~algorithm in
itself which is sometimes called a ``cycle''. We end up with a nested
algorithm where cycles are embedded in a higher-level iterative
scheme.

Furthermore, how to define the $Q_i$'s?  From section~\ref{sec:em}, we
know that the standard~EM auxiliary function $Q(\theta|\theta')$ is
built from the complete-data space~$Z$; see in particular
equation~(\ref{eq:cecl}). Fessler \& Herro introduce {\em hidden-data
spaces}, a concept that generalizes complete-data spaces in the sense
that hidden-data spaces may be coordinate-specific, i.e. there is a
hidden variable~$Z_i$ for each block~$t_i$. Formally, given a block
decomposition $\theta = (t_1, t_2, \ldots, t_s)$, $Z_i$ is a
hidden-data space for $t_i$ if:
$$
p(z_i,y|\theta) = p(y|z_i,\{t_{j\not=i}\}) \, p(z_i|\theta)
$$

This definition's main feature is that the conditional probability
of~$Y$ knowing~$Z_i$ is allowed to be dependent on every parameter
block but~$t_i$. Let us check that the resulting auxiliary function
fulfils the monotonicity condition. We define:
$$
Q_i(\theta|\theta') \equiv E\big[\log p(Z_i|\theta)|y,\theta'\big]
$$

Then, applying Jensen's inequality~(\ref{eq:jensen}) like in
section~\ref{sec:jensen}, we get:
$$
L (\theta) - L (\theta') 
\geq 
Q_i(\theta|\theta') - Q_i(\theta'|\theta')
+ 
\int \log \frac{p(y|z_i,\theta)}{p(y|z_i,\theta')}\,p(z_i|y,\theta')\,dz_i
$$

When~$\theta$ and~$\theta'$ differ only by~$t_i$, the integral
vanishes because the conditional pdf~$p(y|z_i,\theta)$ is independent
from~$t_i$ by the above definition. Consequently, maximizing
$Q_i(\theta|\theta')$ with respect to $t_i$ only (the other parameters
being held fixed) is guaranteed to increase the incomplete-data
likelihood. Specific applications of~SAGE to the Poisson imaging model
or penalized least-squares regression were reported to converge much
faster than standard~EM.

\subsection{CEMM}
\label{sec:cemm} 

Component-wise EM for Mixtures \cite{Celeux-01}. Celeux et al extend
the SAGE methodology to the case of constrained likelihood
maximization, which arises typically in mixture problems where the sum
of mixing proportions should equate to one. Using a Lagrangian
dualization approach, they recast the initial problem into
unconstrained maximization by defining an appropriate penalized
log-likelihood function. The~CEMM algorithm they derive is a natural
coordinatewise variant of~EM whose convergence to a stationary point
of the likelihood is established under mild regularity conditions.

\subsection{AECM}
\label{sec:aecm}

Alternating ECM \cite{Meng-97,Meng-99}. In an attempt to summarize
earlier contributions, Meng \& van Dyk propose to cast a number of
EM~extensions into a unified framework, the so-called~AECM
algorithm. Essentially, AECM~is a SAGE~algorithm (itself a
generalization of both ECM and ECME) that includes another data
augmentation trick. The idea is to consider a family of complete-data
spaces indexed by a working parameter~$\alpha$. More formally, we
define a joint~pdf $q(z,y|\theta,\alpha)$ as depending on
both~$\theta$ and~$\alpha$, yet imposing the constraint that the
corresponding marginal incomplete-data pdf be preserved:
$$
p(y|\theta) = \int q(z,y|\theta,\alpha)\,dz,
$$ and thus independent from~$\alpha$. In other words, $\alpha$~is
identifiable only given the complete data.  A simple way of achieving
such data augmentation is to define~$Z=f_{\theta,\alpha}(Z_0)$,
where~$Z_0$ is some reference complete-data space
and~$f_{\theta,\alpha}$ is a one-to-one mapping for
any~$(\theta,\alpha)$. Interestingly, it can be seen that~$\alpha$ has
no effect if~$f_{\theta,\alpha}$ is insensitive to~$\theta$. In~AECM,
$\alpha$ is tuned beforehand so as that to minimize the fraction of
missing data~(\ref{eq:speed_matrix}), thereby maximizing the
algorithm's global speed. In general, however, this initial
mimimization cannot be performed exactly since the global speed may
depend on the unknown maximum likelihood parameter.

\subsection{PX-EM}
\label{sec:px-em}

Parameter-Expanded EM \cite{Liu-98,Liu-03}. Liu et al revisit the
working parameter method suggested by Meng and van Dyk~\cite{Meng-97}
(see section~\ref{sec:aecm}) from a slighlty different angle. In their
strategy, the joint pdf~$q(z,y|\theta, \alpha)$ is defined so as to
meet the two following requirements. First, the baseline model is
embedded in the expanded model in the sense that $q(z,y|\theta,
\alpha_0)=p(z,y|\theta)$ for some null value~$\alpha_0$. Second, which
is the main difference with~AECM, the observed-data marginals are
consistent up to a many-to-one reduction function~$r(\theta,\alpha)$,
$$ 
p\big(y|r(\theta,\alpha)\big) = \int q(z,y|\theta,\alpha)\,dz,
$$ for all~$(\theta,\alpha)$. From there, the trick is to to
``pretend'' estimating~$\alpha$ iteratively rather than pre-processing
its value.

The PX-EM algorithm is simply an~EM on the expanded model with
additional instructions after the M-step to apply the reduction
function and reset~$\alpha$ to its null value. Thus, given a current
estimate~$\theta_n$, the E-step forms the auxiliary function
corresponding to the expanded model from $(\theta_n,\alpha_0)$, which
amounts to the standard E-step because~$\alpha=\alpha_0$. The M-step
then provides $(\theta^*,\alpha^*)$ such that
$q(y|\theta^*,\alpha^*)\geq q(y|\theta_n,\alpha_0)$, and the
additional reduction step updates $\theta_n$ according to
$\theta_{n+1}=r(\theta^*,\alpha^*)$, implying
$p(y|\theta_{n+1})=q(y|\theta^*,\alpha^*)$. Because
$q(y|\theta_n,\alpha_0)=p(y|\theta_n)$ by construction of the expanded
model, we conclude that $p(y|\theta_{n+1})\geq p(y|\theta_n)$, which
shows that PX-EM preserves the monotonicity property of~EM.

In some way, PX-EM capitalizes on the fact that a large deviation
between the estimate of~$\alpha$ and its known value~$\alpha_0$ is an
indication that the parameter of interest~$\theta$ is poorly
estimated.  Hence, PX-EM adjusts the M-step for this deviation via the
reduction function. A variety of examples where PX-EM converges much
faster than~EM is reported in \cite{Liu-98}. Possible variants of
PX-EM include the coordinatewise extensions underlying~SAGE.

\subsection{Incremental EM}

Following the maximization-maximization approach discussed in
section~\ref{sec:global}, Neal \& Hinton~\cite{Neal-98} address the
common case where observations are i.i.d. Then, we have
$p(y|\theta)=\prod_i p(y_i|\theta)$ and, similarly, the global
EM~objective function~(\ref{eq:maxmax}) reduces to:
$$
L(\theta,q) = 
\sum_i
\big\{ \expt_{q_i} \big[\log p(Z_i,y_i|\theta)\big] 
+
H(q_i)
\big\},
$$ where we can search for~$q$ under the factored form $q(z)=\prod_i
q_i(z)$.  Therefore, for a given~$\theta$, maximizing $L(\theta,q)$
wrt~$q$ is equivalent to maximizing the contribution of each data item
wrt~$q_i$, hence splitting the global maximization problem into a
number of simpler maximizations. Incremental EM~variants are justified
from this remark, the general idea being to update~$\theta$ by
visiting the data items sequencially rather than from a global
E-step. Neal \& Hinton demonstrate an incremental~EM variant for
mixtures that converges twice as fast as standard~EM.

\subsection{Sparse EM}

Another suggestion of Neal \& Hinton~\cite{Neal-98} is to track the
auxiliary distribution $q(z)$ in a subspace of the original search
space~${\cal Q}$ (at least for a certain number of iterations). This
general strategy includes sparse EM~variants where $q$~is updated only
at pre-defined plausible unobserved values. Alternatively,
``winner-take-all'' EM~variants such as the
CEM~algorithm~\cite{Celeux-92} (see section~\ref{sec:cem}) may be seen
in this light.  Such procedures may have strong computational
advantages but, in counterpart, are prone to estimation bias. In the
maximization-maximization interpretation of~EM, this comes as no
surprise since these approaches ``project'' the estimate on a reduced
search space that may not contain the maximum likelihood solution.

\section{Stochastic EM variants}
\label{sec:stochastic_variants}

While deterministic EM~variants were mainly motivated by convergence
speed considerations, stochastic variants are more concerned with
other limitations of standard~EM. One is that the EM auxiliary
function~(\ref{eq:auxiliary}) involves computing an integral that may
not be tractable in some situations.  The idea is then to replace this
tedious computation with a stochastic simulation.  As a typical side
effect of such an approach, the modified algorithm inherits a lesser
tendancy to getting trapped in local maxima, yielding improved global
convergence properties.

\subsection{SEM}

Stochastic EM \cite{Celeux-85}. As noted in
section~\ref{sec:conditional_expectation}, the standard EM auxiliary
function is the best estimate of the complete-data log-likelihood in
the sense of the conditional mean squared error.  The idea underlying
SEM, like other stochastic EM variants, is that there might be no need
to ask for such a ``good'' estimate. Therefore, SEM replaces the
standard auxiliary function with:
$$
\hat{Q}(\theta|\theta') = \log p(z'|\theta'),
$$ where $z'$ is a random sample drawn from the posterior distribution
of the unobserved variable\footnote{Notice that when $Z$ is defined as
$Z=(X,Y)$, this simulation reduces to a random draw of the missing
data~$X$.}, $p(z|y,\theta')$. This leads to the following modified
iteration; given a current estimate $\theta_n$:
\begin{itemize}
\item Simulation step: compute $p(z|y,\theta_n)$ and draw an
unobserved sample $z_n$ from $p(z|y,\theta_n)$.
\item Maximization step: find $\displaystyle \theta_{n+1} =
\arg\max_\theta p(z_n|\theta)$.
\end{itemize}

By construction, the resulting sequence $\theta_n$ is an homogeneous
Markov chain\footnote{The draws need to be mutually independent
conditional to $(\theta_1,\theta_2,\ldots,\theta_n)$, i.e.
$p(z_1,z_2,\ldots,z_n|\theta_1,\theta_2,\ldots,\theta_n)=\prod_i
p(z_i|\theta_i)$.}  which, under mild regularity conditions, converges
to a stationary pdf.  This means in particular that $\theta_n$ doesn't
converge to a unique value!  Various schemes can be used to derive a
pointwise limit, such as averaging the estimates over iterations once
stationarity has been reached (see also SAEM regarding this issue). It
was established in some specific cases that the stationary pdf
concentrates around the likelihood maximizer with a variance inversly
proportional to the sample size. However, in cases where several local
maximizers exist, one may expect a multimodal behavior.

\subsection{DA}

Data Augmentation algorithm \cite{Tanner-87}. Going further into the
world of random samples, one may consider replacing the M-step in SEM
with yet another random draw. In a Bayesian context, maximizing
$p(z_n|\theta)$ wrt~$\theta$ may be thought of as computing the mode
of the posterior distribution $p(\theta|z_n)$, given by:
$$
p(\theta|z_n)
= 
\frac{p(z_n|\theta)p(\theta)}{\int p(z_n|\theta')p(\theta')\, d\theta'}
$$ where we can assume a flat (or non-informative) prior distribution
for~$\theta$.  In DA, this maximization is replaced with a random draw
$\theta_{n+1} \sim p(\theta|z_n)$. From
equation~(\ref{eq:complete_data_space}), we easily check that
$p(\theta|z_n)=p(\theta|z_n,y)$.  Therefore, DA alternates conditional
draws $z_n|(\theta_n,y)$ and $\theta_{n+1}|(z_n,y)$, which is the very
principle of a Gibbs sampler. Results from Gibbs sampling theory
apply, and it is shown under general conditions that the
sequence~$\theta_n$ is a Markov chain that converges in distribution
towards~$p(\theta|y)$. Once the sequence has reached stationarity,
averaging~$\theta_n$ over iterations yields a random variable that
converges to the conditional mean~$\expt(\theta|y)$, which is an
estimator of~$\theta$ generally different from the maximum likelihood
but not necessarily worse.

Interesting enough, several variants of~DA have been proposed recently
\cite{Liu-99,Liu-03} following the parameter expansion strategy
underlying the PX-EM algorithm described in section~\ref{sec:px-em}.

\subsection{SAEM}
\label{sec:saem}

Stochastic Approximation type EM \cite{Celeux-95}. The SAEM algorithm
is a simple hybridation of EM and SEM that provides a pointwise
convergence as opposed to the erratic behavior of SEM.  Given a
current estimate $\theta_n$, SAEM performs a standard EM iteration in
addition to the SEM iteration. The parameter is then updated as a
weighted mean of both contributions, yielding:
$$ 
\theta_{n+1} = 
(1-\gamma_{n+1})\theta^{EM}_{n+1} +
\gamma_{n+1}\theta^{SEM} _{n+1},
$$ where $0\leq\gamma_n\leq 1$. Of course, to apply SAEM, the standard
EM needs to be tractable.

The sequence $\gamma_n$ is typically chosen so as to decrease from~$1$
to~$0$, in such a way that the algorithm is equivalent to SEM in the
early iterations, and then becomes more similar to EM. It is
established that SAEM converges almost surely towards a local
likelihood maximizer (thus avoiding saddle points) under the
assumption that $\gamma_n$ decreases to~$0$ with $\lim_{n\to\infty}
(\gamma_n/\gamma_{n+1})=1$ and $\sum_n \gamma_n \to \infty$.

\subsection{MCEM}

Monte Carlo EM \cite{Wei-90}. At least formally, MCEM turns out to be
a generalization of SEM. In the SEM simulation step, draw
$m$~independent samples $z_n^{(1)}, z_n^{(2)}, \ldots, z_n^{(m)}$
instead of just one, and then maximize the following function:
$$
\hat{Q} (\theta|\theta_n) = \frac{1}{m} \sum_{j=1}^m \log p(z_n^{(j)}|\theta),
$$ 
which, in general, converges almost surely to the standard~EM
auxiliary function thanks to the law of large numbers.

Choosing a large value for~$m$ justifies calling this Monte Carlo
something. In this case, $\hat{Q}$~may be seen as an empirical
approximation of the standard~EM auxiliary function, and the algorithm
is expected to behave similarly to~EM. On the other hand, choosing a
small value for~$m$ is not forbidden, if not advised (in particular,
for computational reasons). We notice that, for~$m=1$, MCEM reduces to
SEM. A possible strategy consists of increasing progressively the
parameter~$m$, yielding a ``simulated annealing'' MCEM which is close
in spirit to~SAEM.

\subsection{SAEM2}

Stochastic Approximation EM \cite{Delyon-99}. Delyon et al propose a
generalization of~MCEM called SAEM, not to be confused with the
earlier~SAEM algorithm presented in section~\ref{sec:saem}, although
both algorithms promote a similar simulated annealing philosophy. In
this version, the auxiliary function is defined recursively by
averaging a Monte Carlo approximation with the auxiliary function
computed in the previous step:
$$ 
\hat{Q}_n(\theta) = 
(1-\gamma_n)\hat{Q}_{n-1}(\theta)
+ \frac{\gamma_n}{m_n} \sum_{j=1}^{m_n} \log p(z_n^{(j)}|\theta),
$$ where $z_n^{(1)}, z_n^{(2)}, \ldots, z_n^{(m_n)}$ are drawn
independently from~$p(z|y,\theta_n)$. The weights~$\gamma_n$ are
typically decreased across iterations in such a way
that~$\hat{Q}_n(\theta)$ eventually stabilizes at some point. One may
either increase the number of random draws~$m_n$, or set a constant
value~$m_n\equiv 1$ when simulations have heavy computanional cost
compared to the maximization step. The convergence of~SAEM2 towards a
local likelihood maximizer is proved in \cite{Delyon-99} under quite
general conditions.

Kuhn et al~\cite{Kuhn-02} further extend the technique to make it
possible to perform the simulation under a
distribution~$\Pi_{\theta_n}(z)$ simpler to deal with than the
posterior pdf~$p(z|y,\theta_n)$. Such a distribution may be defined as
the transition probability of a Markov chain generated by a
Metropolis-Hastings algorithm. If~$\Pi_\theta(z)$ is such that its
associated Markov chain converges to~$p(z|y,\theta)$, then the
convergence properties of~SAEM2 generalize under mild additional
assumptions.

\section{Conclusion}

This report's primary goal is to give a flavor of the current state of
the art on EM-type statistical learning procedures.  We also hope it
will help researchers and developers in finding the literature
relevant to their current existential questions. For a more
comprehensive overview, we advise some good tutorials that are found
on the
internet~\cite{Couvreur-96,Bilmes-98,BergerA-98,Dellaert-02,vanDyk-00,Liu-03}.

\appendix

\section{Appendix}

\subsection{Maximum likelihood quickie}
\label{app:ml}

Let~$Y$ a random variable with pdf $p(y|\theta)$, where $\theta$ is an
unknown parameter vector. Given an outcome~$y$ of~$Y$, maximum
likelihood estimation consists of finding the value of~$\theta$ that
maximizes the probability~$p(y|\theta)$ over a given search
space~$\Theta$. In this context, $p(y|\theta)$ is seen as a function
of~$\theta$ and called the likelihood function. Since it is often more
convenient to manipulate the logarithm of this expression, we will
focus on the equivalent problem of maximizing the log-likelihood
function:
$$
\hat{\theta}(y) = \arg\max_{\theta\in\Theta} L(y,\theta)
$$ where the log-likelihood $L(y,\theta) \equiv \log p(y|\theta)$ is
denoted $L(y,\theta)$ to emphasize the dependance in $y$, contrary to
the notation $L(\theta)$ usually employed throghout this report.
Whenever the log-likelihood is differentiable wrt $\theta$, we also
define the score function as the log-likelihood gradient:
$$
S(y,\theta) = \frac{\partial L}{\partial \theta}(y,\theta)
$$

In this case, a fundamental result is that, for all vector
$U(y,\theta)$, we have:
$$
\expt(SU^t) = \frac{\partial}{\partial \theta} \expt(U^t) 
- \expt \Big( \frac{\partial U^t}{\partial \theta} \Big)
$$ where the expectation is taken wrt the distribution
$p(y|\theta)$. This equality is easily obtained from the logarithm
differentiation formula and some additional manipulations. Assigning
the ``true'' value of $\theta$ in this expression leads to the
following:
\begin{itemize}
\item $\expt(S)=0$
\item $\displaystyle \cov(S,S)=- \expt \Big( \frac{\partial S^t}{\partial
\theta} \Big)$ (Fisher information)
\item If $U(y)$ is an unbiased estimator of $\theta$, then
$\cov(S,U)={\rm Id}$.
\item In the case of a single parameter, the above result implies
$\var(U) \geq \frac{1}{\var(S)}$ from the Cauchy-Schwartz inequality,
i.e.  the Fisher information is a lower bound for the variance
of~$U$. Equality occurs iff $U$ is an affine function of $S$, which
imposes a specific form to $p(y|\theta)$ (Darmois theorem).
\end{itemize}

\subsection{Jensen's inequality}
\label{app:jensen}

For any random variable~$X$ and any real continuous concave
function~$f$, we have:
$$
f\big[\expt(X)\big] \geq \expt\big[f(X)\big],
$$

If $f$ is strictly concave, equality occurs iff $X$ is deterministic.

%\bibliography{stat} 
%\bibliographystyle{abbrv}

\end{document}